\documentclass[unnumsec,webpdf,contemporary,large,namedate, biblatex]{oup-authoring-template}

\usepackage{graphicx}
\graphicspath{{figures/}}
\usepackage{booktabs} 
\usepackage{makecell} 
\usepackage{amsmath}  
\usepackage{float} 

\theoremstyle{thmstyleone}%
\theoremstyle{thmstyletwo}%
\theoremstyle{thmstylethree}%

\usepackage{booktabs,tabularx,pifont}

\usepackage{tabularx,booktabs,array,ragged2e}
\newcolumntype{L}[1]{>{\RaggedRight\arraybackslash}p{#1}}
\newcolumntype{Y}{>{\RaggedRight\arraybackslash}X}
\newcommand{\twolineskip}{\par\vspace{0.5\baselineskip}}

\begin{document}
\setlength{\parindent}{00pt}

\journaltitle{Bioinformatics}
\DOI{DOI HERE}
\copyrightyear{2026}
\pubyear{2026}
\access{Advance Access Publication Date: Day Month Year}
\appnotes{Application Note}

\firstpage{1}


\title[AF\_Cache]{AF\_Cache: Efficient Pipeline for Running AlphaFold for High-Throughput Protein-Protein Interaction Prediction}
 
\author[1]{Sarah Narrowe\ORCID{0000-0003-0087-8756}} 
\author[1]{Arne Elofsson\ORCID{0000-0002-7115-9751}}
\author[2$\ast$]{Claudio Mirabello\ORCID{0000-0002-7115-9751}}

\authormark{Narrowe et al.}

\address[1]{\orgdiv{Department of Biochemistry and Biophysics and SciLifeLab}, \orgname{Stockholm University}, \orgaddress{\street{Box 1031}, \postcode{171 21}, \state{Stockholm}, \country{Sweden}}}
\address[2]{\orgdiv{Dept of Physics, Chemistry and Biology, National Bioinformatics Infrastructure Sweden, Science for Life Laboratory}, \orgname{Linköping University}, \orgaddress{\postcode{581 83}, \state{Linköping}, \country{Sweden}}}
\corresp[$\ast$]{Corresponding author. \href{email:email-id.com}{claudio.mirabello@scilifelab.se}}




\abstract{\textbf{Motivation:} Accurate prediction of protein-protein interactions is essential for understanding biological processes, and recent advances such as AlphaFold2 and AlphaFold3 have enabled structure-based interaction prediction at unprecedented accuracy. However, the high computational cost of these methods, driven primarily by CPU-based repeated multiple sequence alignment (MSA) generation and, for AlphaFold2, repeated model recompilations, limits their applicability in large-scale, high-throughput settings. This creates a need for efficient pipelines that retain predictive performance while substantially reducing runtime.\\\\
\textbf{Results}: We present AF\_Cache, a high-throughput Nextflow pipeline for accelerating protein-protein interaction prediction using AlphaFold2 and AlphaFold3. AF\_Cache combines GPU-accelerated MSA generation with MMseqs2, feature caching to eliminate redundant alignment computations, and sequence length bucketing to minimise repeated JAX compilations. Benchmarking on a dataset of 5,050 human mitochondrial protein pairs demonstrates a $\sim$2-fold reduction in inference time for AlphaFold2 and up to a 13-fold speedup of the MSA generation.  AF\_Cache enables efficient large-scale interaction screening and provides a practical framework for deploying AlphaFold-based methods in high-throughput applications.\\
\textbf{Availability and implementation:}  The code and Nextflow pipeline are available on GitHub here: \url{https://github.com/clami66/AF_cache}. The code for reproducing the results of the paper, the MSAs, and the predicted models can be found at  Zenodo: \url{https://zenodo.org/records/20478892}. \\
\textbf{Supplementary information}: Supplementary information is available online.\\
}


\maketitle

\section{Introduction}\label{introduction}
\begin{figure}
\centering
\includegraphics[width=0.5\textwidth]{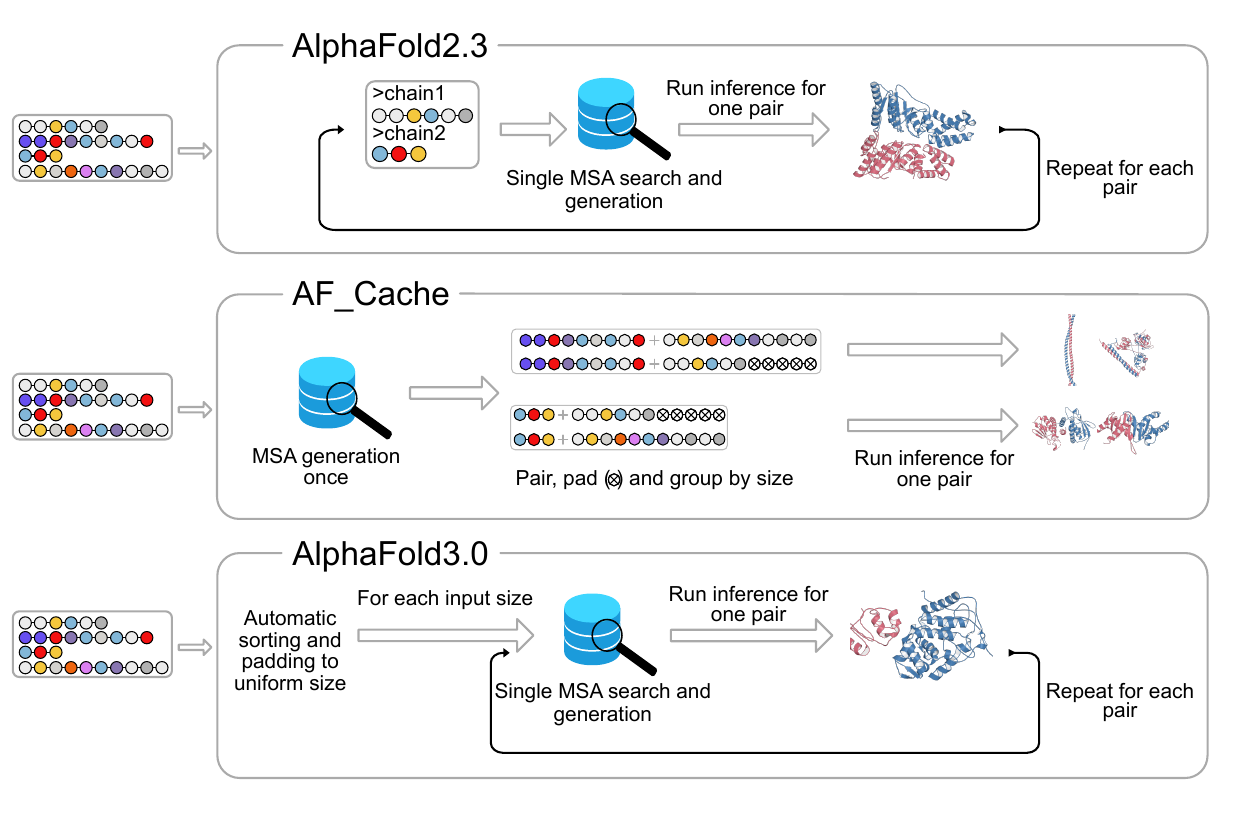}
\caption{\textbf{Overview of the AF\_Cache and AlphaFold pipelines.} The input to the pipeline is a directory containing the input FASTA files. It is possible to either run all-against-all (the default) or use an input file. The next step of the pipeline is to generate MSAs for all proteins simultaneously using MMseqs2-GPU. Then the pairs are divided into size buckets and padded to a common size. Next, each pair is executed sequentially, with JAX compilation occurring only for the first pair in each size bucket. There is a maximum number of pairs to be executed for each size; multiple runs of each size are allowed. For both versions of AlphaFold, the default way to generate MSAs for each chain is right before inference, thus not reusing the MSAs for the same chains when predicting different pairs. AF3 automatically employs padding of different bucket sizes for both monomers and multimers, whereas AF2 implements this only for monomers. AF2 uses only FASTA files as input, whereas AF3 accepts a JSON file or a directory of JSON files. For the default version of AF3, the user has to write their own JSON files, while this is handled automatically in AF\_Cache.}
\label{fig:af_cache_pipeline}
\end{figure}

Knowing which proteins interact with each other is crucial for understanding biological processes. The release of AlphaFold2 (AF2)\citep{alphafold2,folddock} enabled the prediction and identification of protein-protein interactions with unprecedented accuracy. However, for large-scale interaction predictions, the default AF2/3 pipelines are often too slow. There are two major factors that contribute to AF2's high inference times, and one of them also affects AlphaFold3 (AF3)\citep{af3}: MSA generation and JAX compilation. The multiple-sequence alignment (MSA) generation step of AlphaFold uses CPU-based JackHMMer \citep{jackhmmer} and HHblits \citep{HHblits}. However, faster MSA methods exist \citep{mmseqs}. Fast batched MSA generation has been used previously by Perry et al. in~\citep{AlphaFast} and is also used in ColabFold\citep{colabfold}. Here, we propose an alternative method that introduces further speedups and also supports AF2.
\twolineskip
The second factor contributing to the time is that JAX neural network models must be recompiled for sequences of varying lengths. AF2 provides an option to pad sequences to equal lengths, but it is only available for monomers and is not divided into multiple size buckets, making it highly inefficient for shorter sequences. The JAX compilation time is negligible for large proteins. However, for smaller proteins, it can be the most time-consuming part of the predictions. 
\twolineskip
Here, we present AF\_Cache, which addresses both shortcomings. Figure \ref{fig:af_cache_pipeline} provides an overview. Using MMseqs2-GPU together with padding protein pairs within size buckets, we achieve a 13x speedup in MSA generation and a 2x speedup in protein folding (inference/prediction) for AF2.  The corresponding speed improvement for AF3 is 5x for MSA generation, while utilising the same inference code as the default version, i.e. there is no speedup there. We provide the AF\_Cache pipeline in a ready-to-use Nextflow pipeline for the latest versions of AF2 and AF3. The pipeline automatically downloads and installs dependencies, including databases and network parameters (AF2 only), whenever necessary, and is an easy way to install AF2 or AF3 on local and HPC systems.

\section{Methods}
AF\_Cache implements several optimisations to improve input feature generation and model compilation efficiency (the latter applies only to AF2). 

\subsection{MSA Generation}\label{subsec:msa-gen}
MSAs for a dataset are generated in a single step with MMseqs2, which considerably reduces runtime compared to generating MSAs per sequence. We furthermore use the GPU-accelerated version of MMseqs2, in which sequence and profile searches are performed on a GPU. A single-GPU MMseqs2 run is faster than a CPU-only run, and the speedup increases with multiple GPUs. In the benchmarks reported here, MMseqs2 was run on a single GPU.
\twolineskip
To generate MSAs for AF2 and AF3, we follow the same search and alignment protocol as ColabFold \citep{colabfold}, with one additional improvement: when performing alignments on both UniRef and the environmental databases, CPU and GPU steps are run in parallel across the two databases. This increases efficiency by reducing the time the GPU idles while waiting for the UniRef alignment step to complete.

\subsection{Generation of Input Features}
When making predictions for sets of multimers, AlphaFold will repeatedly run the alignment steps. This means that whenever two or more target complexes share one or more monomer sequences, duplicate alignment steps will be performed, wasting CPU resources. For $N$ proteins screened as all unique dimers including homodimers, default per-chain MSA generation requires $N(N+1)$ chain-level MSA generations, whereas only $N$ unique monomer MSAs are needed. Here, we implement automatic caching of MSAs and template features so that, for a given set of unique protein monomers, all alignments and templates are generated only once ($N$ alignment steps are performed). Features are stored as \emph{pickle} files in the AlphaFold2 version of the pipeline and as \emph{JSON} input files in the AlphaFold3 version.

\subsection{Model Compilation Optimization}
In AF2, we implement bucketing of multimers of similar size (defined as the sum of the lengths of protein chains). We perform inference once on a bucket of multimers while padding all feature tensors to the same protein length. This means the neural network model will be compiled only once for the first protein multimer, saving 1-2 minutes of GPU time for each additional multimer during inference. This feature is already available in AF3, so we did not implement it there.

\subsection{Proteins Used for Benchmarking}\label{subsec:data-used}
The proteins used in the analysis were retrieved on the 29th of January 2026 from the Human Protein Atlas (HPA) through \url{proteinatlas.org} \citep{hpa-subcellular-info}. The proteins were limited to mitochondrial proteins identified with the filters \emph{subcell\_location:Mitochondria AND hpa\_evidence:Evidence at protein level}, resulting in a total of 821 proteins. To narrow it down to 100 proteins, we randomly sampled 100 of these between 40 and 1000 residues. The protein sequences were retrieved from UniProt \citep{uniref30}. The UniProt IDs, sequence lengths and sequences of all 100 proteins are available in the Supplementary CSV file. PPI predictions were performed all-against-all for the 100 selected proteins, including homodimeric interactions, using AF\_Cache2/3, AlphaFold2.3, and AlphaFold3.0. Symmetric pairs are excluded by default, meaning that for two proteins A and B, only A-B is included and not B-A; these can optionally be included using the \textit{-{}-both\_directions} flag. This resulted in 5,050 unique pairs: $100 \times 99 / 2 = 4,950$ heterodimers plus 100 homodimers. 
\twolineskip
To identify highly similar PDB entries for each protein, we used MMseqs2 with a fractional sequence identity \emph{(fident)} threshold of 0.7 against PDB. This threshold accounts for the fact that many PDB sequences are shorter than their corresponding UniProt sequences. When two or more proteins are mapped to the same PDB entry, we classify each possible pair as having a ~\emph{shared PDB entry}, as not all pairs with a shared PDB entry interact directly. 

\subsection{AlphaFold Prediction Settings}\label{subsec:default-predictions}
When running AF2 (either in AF\_Cache or default), only one model was used (\textit{model\_1\_multimer\_v3}) to make a single prediction. The maximum number of recycles was set to 3. Template generation was disabled by running AF\_Cache with the \textit{-{}-skip\_templates} option and using a dummy template database in the default version. AF3 was run with a single random seed and one diffusion sample. The MSAs for the default AlphaFold pipelines were generated on a separate CPU cluster.
\twolineskip
The MSAs were generated using the same tools and databases as the original AlphaFold pipelines. These tools were JackHMMer\citep{jackhmmer} and HHBLITS \citep{HHblits}. The databases used were full BFD, small BFD, UniRef30, mgnify, UniProt, and UniRef90 \citep{ uniref30,bfd, mgnify}. The default for AF2 is full BFD, while the default for AF3 is small BFD. 
\twolineskip
The MSA-generation part of the default AlphaFold pipelines was performed on 8, 16, or 32 Intel Xeon Gold 6130 CPU cores (depending on the required memory), while the GPU jobs were run on an NVIDIA A100 with 40 GB of memory.

\subsection{Differences Between AlphaFast and AF\_Cache}
In AlphaFast, \citep{AlphaFast} Perry et al. propose to speed up MSA generation for AlphaFold3 with batched input sequences to MMseqs2-GPU. A key difference in our work is that AF\_Cache generates all monomer MSAs for the full dataset at the start of the workflow, rather than processing them in batches. Furthermore, we parallelise some GPU and CPU tasks during the MSA generation step to improve efficiency. Another major difference is that AlphaFast uses MMseqs2 versions of the same databases as the original AlphaFold3 paper, whereas AF\_Cache uses the ColabFold databases. Finally, our framework includes both AlphaFold2 and AlphaFold3 and allows for parallelisation across multiple nodes on HPC clusters, as well as the ability to run single jobs on local machines. 

\section{Results and Discussion}\label{results}
To demonstrate the speed improvements made by AF\_Cache we used the 5,050 mitochondrial protein pairs for both structure prediction and MSA generation and compared the timings for each method. The ipTM scores for the runs were compared to assess the similarity of the results. 
\begin{figure}
\centering
\includegraphics[width=0.5\textwidth]{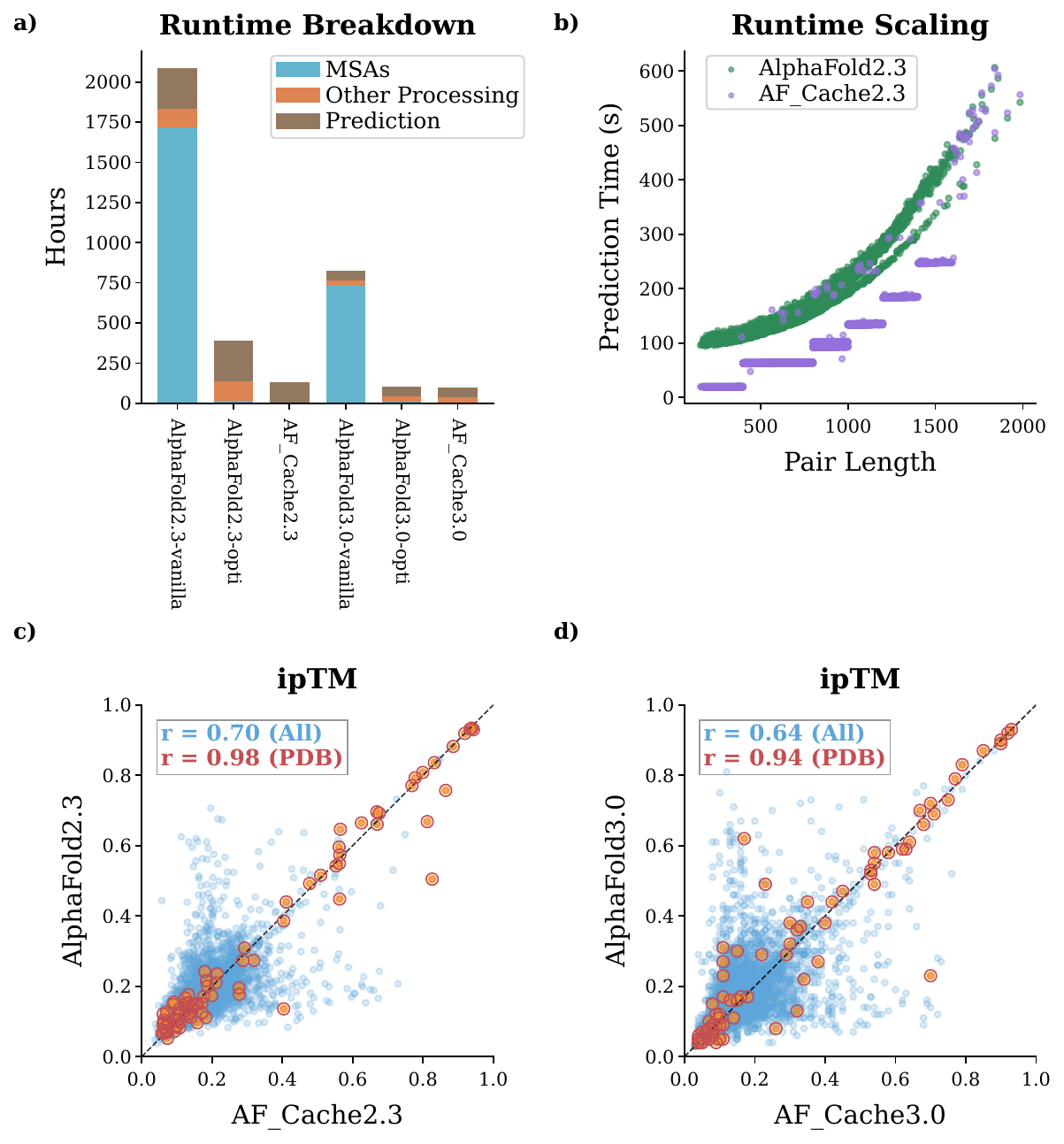}
\caption{\textbf{AF\_Cache speeds up AF2 and AF3 and achieves similar structural scores on pairs with shared PDB entries.} \textbf{a)} Runtime breakdown, here the time spent on MSAs, other prediction tasks and prediction (inference) time is separated for each pipeline. 8, 16, or 32 cores were used to generate the MSAs for the default pipeline, but here we have taken the total core time and divided it by 128, representing 128 cores, since the tasks can easily be parallelised. Other pre-prediction tasks include feature processing (by default) and cache generation for the AF\_Cache pipeline. \textbf{b)} Runtime scaling. Here, the length of the protein pairs vs the prediction time is plotted in a scatter plot. The scaling behaviour is the same for both methods, but AF\_Cache is, on average, 90 seconds faster per pair than the default since it requires fewer JAX model compilations. \textbf{c-d)} Comparison of ipTM scoring of pairs comparing default AlphaFold versions and the cache versions. Pearson correlation of $r=0.64-0.70$ across all pairs and $r \ge 0.94$ for pairs with shared PDB entries (see methods).} 
\label{fig:runtime-performance-comparison}
\end{figure}

\subsection{Pre-Prediction Runtimes}\label{subsec:pre-pred-times}
The first step in both the default and cache pipelines is to generate MSAs, as explained in Section~\ref{subsec:msa-gen}.   
\twolineskip
Using MMseqs2-GPU, there is a 1702-fold speedup in raw GPU-to-CPU core hours with the full BFD, and a 688-fold decrease with the small BFD. However, this does not represent a real-world scenario since CPU cores are considerably cheaper than GPUs. Therefore, the core hours in Figure \ref{fig:runtime-performance-comparison}A are rescaled to 128 cores, assuming perfect parallelisation. When comparing this more realistic scenario, the speedups are reduced to 13-fold for the full BFD and 5-fold for the smaller database, still a considerable increase for high-throughput applications.  
\twolineskip
Due to restrictions on our high-performance computing (HPC) cluster, the CPU and GPU components of the AlphaFold default pipelines had to be run separately. This could have led to different timings compared to running the true default AlphaFold pipelines. Additionally, the default AlphaFold pipelines partition the database, leading to more efficient memory usage and better parallelisation than our MSA generation does. By rescaling the comparisons in this study to 128 cores, we partially compensate for this imbalance but cannot guarantee complete compensation. 
\twolineskip
Instead of regenerating every MSA for every chain in every pair, the MSAs were generated once for each of the 100 proteins with symbolic links tying them to the correct directory for each protein pair (AF2) or the MSAs were included in the input\_jsons (AF3). This approach is called AlphaFold2/3-opti, and the original approach with repeated MSA generation for each chain is called AlphaFold2/3-vanilla, as shown in Figure \ref{fig:runtime-performance-comparison}A. Since the all-against-all benchmark contains 5,050 dimers, corresponding to 10,100 chain occurrences, the vanilla MSA runtime was estimated as 101 times the runtime required to generate MSAs once for the 100 unique proteins. When comparing AF\_Cache to the vanilla versions of AlphaFold, the speedup increases to 1343-fold (AF2) and 542-fold (AF3), still rescaled to 128 cores. A more detailed breakdown of the exact hours, as well as the runtime in seconds per target, can be found in the supplementary tables \ref{supp:tab:runtime-breakdown-table} and \ref{supp:tab:runtime-breakdown-table-per-target}.
\twolineskip
The final part of the cache pre-prediction pipeline is the \emph{parse\_features} step. This generates caches for the AF2 implementation and templates (which can be skipped with the \textit{-{}-skip\_templates} option). This part takes an average of 20 seconds per protein. For the AF3 pipeline, the average is 10 seconds. The average time is shorter because caching is not needed; it is handled automatically by AF3. 

\subsection{Prediction Runtimes}\label{subsec:pred-times}
When comparing prediction runtimes between the Cache pipelines and the default versions, it only makes sense to compare the AF2 versions, since there is no difference in the prediction stage between AF3 and AF\_Cache3.0. However, an important note: the AF3 version affects the model's runtime. During internal testing, we noticed that the latest version, 3.0.1, is notably faster than 3.0.0; the pipeline automatically uses 3.0.1. 
\twolineskip
Implementing caching and padding sequences to equal lengths reduced the total prediction and compilation time from 253 to 125 GPU hours, a reduction of more than 50\%. Figure \ref{fig:runtime-performance-comparison}B clearly shows that the scaling between the two implementations is identical, but that AF\_Cache is on average 90 seconds faster per protein pair than the default version of AF2. When comparing Figure \ref{fig:runtime-performance-comparison}B and Supplementary Figure \ref{supp:fig:af3-scaling-behavior}, both the default and AF\_Cache version of AF2 are noticeably slower than AF3. Notably, in Figure \ref{fig:runtime-performance-comparison}B, the prediction times appear to follow two curves from around 1200 residues in pair length. This is most likely due to the fact that AlphaFold employs an \textit{early-stop-tolerance} parameter that will only run an additional recycle if the RMSD between the current predicted structure and the previous cycle's structure exceeds 0.5 Ångström (Å) \citep{ColabFoldMajorChanges}. Subsequently, predicted structures with lower RMSD will stop after 2 recycles, thus shortening the prediction time of those pairs.   

\subsection{ipTM Comparisons}\label{subsec:ipTM-comps}
To exemplify the pipeline's usability, in addition to the speedup, we show in Figure \ref{fig:runtime-performance-comparison}C-D that the correlations across all pairs were moderate, whereas correlations among pairs with shared PDB-entry support were high. The Pearson Coefficient is 0.64 for AF3 and 0.70 for AF2 (P-values less than $10^{-300}$). This moderate correlation is consistent with previous findings from Todor et al\citep{pooled-alphafold}. Furthermore, the correlations between ipTM scores in AF2 and AF3 are 0.42 for AF\_cache and 0.41 for the default implementation (see Supplementary Figures \ref{supp:fig:afcache23vs30} and \ref{supp:fig:default23vs30_default}). Correlation increases to 0.98 for AF2, 0.94 for AF3, and 0.92-0.94 (P-values smaller than $10^{-30}$) between AF2 and AF3 for pairs with shared PDB entries, indicating that pairs sharing a PDB entry are consistently predicted. 

\section{Conclusions}\label{conclusions}

In this work, we introduced AF\_Cache, a high-throughput pipeline that reduces the computational cost of large-scale protein–protein interaction prediction using AlphaFold2 and AlphaFold3. By combining GPU-accelerated MSA generation with MMseqs2, feature caching, and sequence-length bucketing to minimise redundant JAX compilations, we achieve substantial speedups in both pre-processing and inference (for AlphaFold2). These improvements enable efficient all-against-all prediction workflows that would otherwise be computationally expensive using the default AlphaFold pipelines. While we observe moderately similar correlations in prediction scores across different runs and pipeline configurations, as reported previously, we also observe consistently high correlations for structurally supported pairs, indicating that structurally relevant signals are preserved. AF\_Cache provides a practical and scalable solution for large-scale, even proteome-wide, interaction screening.

\section{Code and Data Availability}\label{code_avail}
All code and results are available on GitHub: \url{https://github.com/clami66/AF_cache} and Zenodo: \url{https://zenodo.org/records/20478892}.  
\section{Competing interests}
No competing interests are declared.
\section{Author contributions statement}
SN ran all the benchmark experiments, wrote the first draft of the manuscript, and prepared the figures. CM wrote the code for AF\_Cache pipelines, helped with writing and analysis, oversaw the project, and handled all correspondence. AE helped with funding, writing, analysis and supervision. 

\section{Acknowledgments}
The authors thank anonymous reviewers for their valuable suggestions. AE was funded by Vetenskapsrådet, Grant No. 2021-03979, and the Knut and Alice Wallenberg Foundation, Grant No. 2022.0032. CM was financially supported by the SciLifeLab \& Wallenberg Data Driven Life Science Program, Knut and Alice Wallenberg Foundation (KAW 2020.0239, KAW 2017.0003), and by the National Bioinformatics Infrastructure Sweden (NBIS) at SciLifeLab. The computations and data handling were enabled by the supercomputing resource Berzelius, provided by the National Supercomputer Centre at Linköping University, the Knut and Alice Wallenberg Foundation, and SNIC, grant numbers SNIC 2021/5-297 and Berzelius-2021-29. The authors also express their gratitude to the rest of the Arne Elofsson lab for valuable discussions and input, and to Dr. Mahesh Binzer-Panchal at NBIS for valuable input on the Nextflow pipeline.

\bibliographystyle{science}
\bibliography{reference, refs}

\clearpage 
\newpage
\section{Supplementary Material}\label{supplementary_material}
\renewcommand{\thepage}{S\arabic{page}}
\renewcommand{\thesection}{S\arabic{section}}
\renewcommand{\thetable}{S\arabic{table}}
\renewcommand{\thefigure}{S\arabic{figure}}
\setcounter{figure}{0}
\setcounter{table}{0}
\setcounter{page}{1}

\subsection{Table \ref{supp:tab:runtime-breakdown-table}: Runtime Breakdown Table}
\begin{table}[bh]
\centering
\caption{Runtime breakdown for AlphaFold2.3, AlphaFold3.0, and AF\_Cache runs.}
\label{supp:tab:runtime-breakdown-table}
\begin{tabular}{lccc}
\toprule
                 Run &  MSAs (h) &  Other (h) &  Prediction (h) \\
\midrule
AlphaFold2.3 vanilla &   1732.92 &     116.89 &          253.16 \\
   AlphaFold2.3 opti &     17.16 &     116.89 &          253.16 \\
        AF\_Cache2.3 &      1.29 &       2.07 &          125.16 \\
AlphaFold3.0 vanilla &    736.87 &      33.55 &           61.96 \\
   AlphaFold3.0 opti &      7.30 &      33.55 &           61.96 \\
        AF\_Cache3.0 &      1.36 &      33.83 &           61.96 \\
\bottomrule
\end{tabular}
\end{table}

\subsection{Table \ref{supp:tab:runtime-breakdown-table-per-target}: Runtime Breakdown Table - Per Target (protein pair)}
\begin{table}[bh]
\centering
\caption{Runtime breakdown for AlphaFold2.3, AlphaFold3.0, and AF\_Cache runs in seconds per target protein pair.}
\label{supp:tab:runtime-breakdown-table-per-target}
\begin{tabular}{lrrrr}
\toprule
Run & MSAs (s/target) & Other (s/target) & Prediction (s/target) & Total (s/target) \\
\midrule
AlphaFold2.3 vanilla &          1235.3 &             83.3 &                 180.5 & 1499.1 \\
   AlphaFold2.3 opti &            12.2 &             83.3 &                 180.5 &            276.0 \\
        AF\_Cache2.3 &             0.9 &              1.5 &                  89.2 &             91.6 \\
AlphaFold3.0 vanilla &           525.3 &             23.9 &                  44.2 &            593.4 \\
   AlphaFold3.0 opti &             5.2 &             23.9 &                  44.2 &             73.3 \\
        AF\_Cache3.0 &             1.0 &             24.1 &                  44.2 &             69.3 \\
\bottomrule
\end{tabular}
\end{table}

\clearpage
\newpage

\begin{figure}[bh]
\centering
\includegraphics[width=0.5\textwidth]{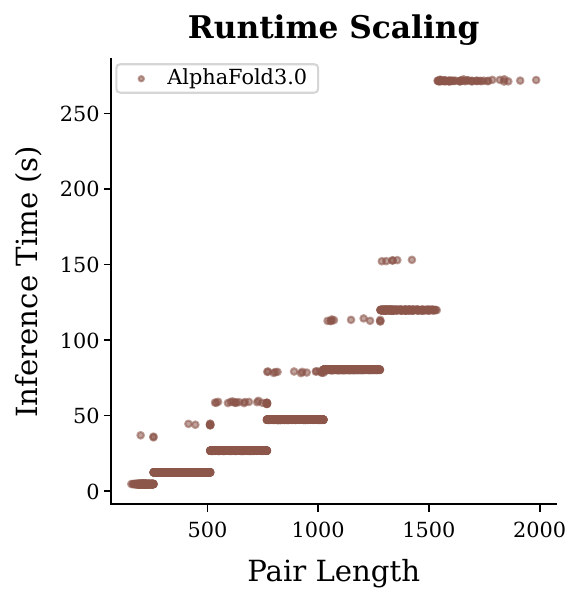}
\caption{Plot of inference time vs pair length for AlphaFold3.}
\label{supp:fig:af3-scaling-behavior}
\end{figure}

\subsection{Figure \ref{supp:fig:af3-scaling-behavior}: AlphaFold3 Scaling Behavior}

\begin{figure}[bh]
\centering
\includegraphics[width=0.5\textwidth]{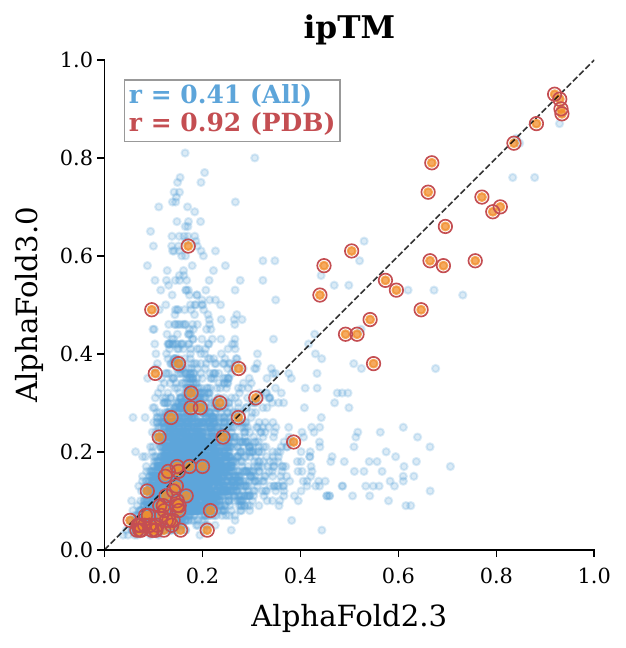}
\caption{Plot of correlation between ipTM values of default AF2.3 and default AF3.0.}
\label{supp:fig:default23vs30_default}
\end{figure}

\subsection{Figure \ref{supp:fig:default23vs30_default}: Default AlphaFold2.3 vs Default AlphaFold3.0}

\begin{figure}[bh]
\centering
\includegraphics[width=0.5\textwidth]{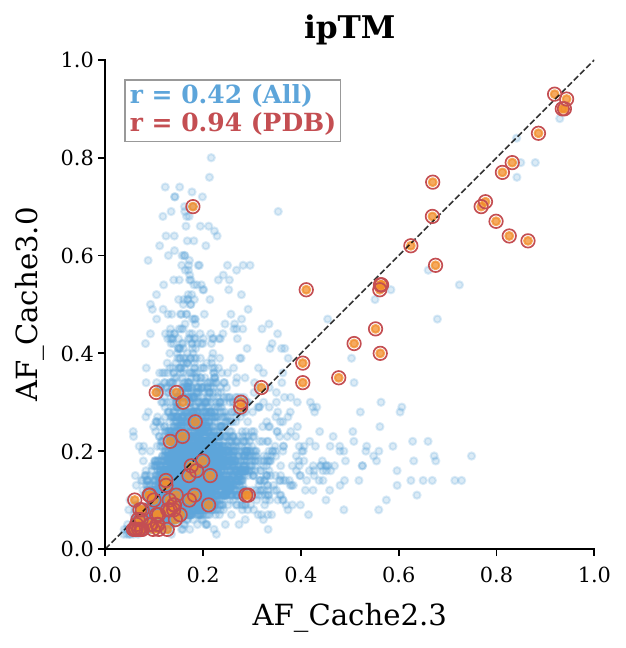}
\caption{Plot of correlation between ipTM values of AF\_Cache2.3 and AF\_Cache3.0.}
\label{supp:fig:afcache23vs30}
\end{figure}

\subsection{Figure \ref{supp:fig:afcache23vs30}: AF\_Cache2.3 vs AF\_Cache3.0}


\end{document}